\documentclass[a4paper,11pt, showpacs, amsmath, amssymb]{revtex4}

\usepackage{color}

\begin{document}

\title{Constructing conformally invariant equations using Weyl geometry}
\author{Sofiane Faci}
\email{sofiane@cbpf.br}
\affiliation{Institute of Cosmology, Relativity and Astrophysics (ICRA - CBPF),
Rua Dr. Xavier Sigaud, 150, Urca, CEP 22290-180, Rio de Janeiro, RJ, Brasil
} 
\date{\today}% It is always \today, today,

\begin{abstract}
We present a simple, systematic and practical method to construct conformally invariant equations in arbitrary Riemann spaces. 
 This method that we call  "Weyl-to-Riemann" is based on two features of Weyl geometry. 
i) A Weyl space is defined by the metric tensor and the Weyl vector $W$, it becomes equivalent to a Riemann space when $W$ is gradient.
ii) Any homogeneous differential equation written in a Weyl space by means of the Weyl connection is conformally invariant. 
The Weyl-to-Riemann method selects those equations whose conformal invariance is preserved when reducing to a Riemann space.
Applications to scalar, vector and spin-2 fields are presented, which demonstrates the efficiency of the present method.
In particular, a new conformally invariant spin-2 field equation is exhibited. This equation extends Grishchuk-Yudin's equation and fixes its limitations since it does not require the Lorenz gauge. Moreover this equation reduces to the Drew-Gegenberg and Deser-Nepomechie equations in respectively  Minkowski and de Sitter spaces.
\end{abstract}

\pacs{11.25.Hf, 11.15.-q, 03.65.Fd, 04.20.Cv}% PACS, the Physics and Astronomy
%11.25.Hf Conformal field theory, algebraic structures
%11.15.-q Gauge field theories
%03.65.Fd Algebraic methods
%04.20.Cv Fundamental problems and general formalism
\maketitle

%\tableofcontents

\section{Introduction}
Conformal symmetry is a fundamental ingredient for theoretical physics. The underlying reason is that conformal transformations preserve causal structures and propagate the modes of  conformally invariant equations on light-cones \cite{Deser:1983}.

The present work deals with the significant problem of how to construct such equations. This is a non-trivial question and several answers are proposed in the literature.
They include Boulanger's method \cite{Boulanger:2004} and the Unfolded Dynamics Approach \cite{Shaynkman:fk, Vasiliev:2007fk, Vasiliev2010176}, both are highly sophisticated but not very practical. 
The same critic applies to the Tractor calculus \cite{Gover:2008-tractors} and the $D+2$ dimensional ambient space techniques \cite{Fefferman:2007}. Other methods are designed to deal exclusively with scalar fields \cite{Manvelyan:2007, Erdmenger:1997}. 
There is also the so-called Ricci-gauging method which was developed by Iorio \textit{et al.} \cite{Iorio:1996}. The authors start with an $SO(2,4)$ invariant equation in Minkowski space and extend it to reach a Weyl invariant equation in arbitrary Riemann spaces. However, despite its beauty and simplicity it is not  a systematic procedure and has been applied only to second-order equations.
\textit{See refs. \cite{ Zumino, pconf7} for the link between Weyl and $SO(2,4)$ transformations}.

%The method
We present a new method that we call  ``Weyl-to-Riemann'' and which takes the opposite direction to that of Iorio \textit{et al.}.  We start from a more general geometry than the Riemann one, namely Weyl geometry. The latter relies on two  objects: the metric tensor and the Weyl vector $W$.
On the one hand, when $W$ is gradient the corresponding Weyl space  is equivalent to a Riemann space \cite{Drechsler:1999fk}.  
On the other hand, using the Weyl covariant derivative and geometrical tensors, defined by means of the Weyl connection, any equation written in a Weyl space is conformally invariant. The only condition is the homogeneity of this equation, which means that its different terms must have the same conformal weight.
In general, this equation depends on $W$ and the game consists in selecting those equations for which $W$ decouples, at least when $W=df$. Also the resulting equation is conformally invariant in a Riemann space.

In practice we look for the right combination of the different terms that cancels $W$.  Imposing $W=df$ then becomes invisible and thereof conformal invariance is preserved at the Riemann stage.
If such a combination does not exist then the residual gradient $df$ mights break the conformal invariance. 
%In this case, $df$ can be interpreted as a mass term in the same spirit as in refs. \cite{Drechsler:1999fk, Gover:2008}. 
In some situations, $W$ does not decouple in a Weyl space but cancels when it has a gradient form or when some extra constraints are imposed and then recover conformal invariance.
The aim of this method is to provide a simple, practical and systematic algorithm. It translates the problem of constructing conformally invariant equations of any order into algebraic systems whose solutions  lead to the desired equations.

The Weyl-toRiemann method was successfully applied to study the conformal properties of scalar, vector and spin-2  fields. In particular,  we exhibit the most general conformally invariant second order equation for a spin-2 field $h_{\mu\nu}$ defined in an arbitrary Riemann space with a canonical weight $+1$. It reads
\begin{equation*}\label{}
\begin{split}
 \Box \, h_{\mu\nu}  - \frac{2}{3} \nabla_{(\mu}\nabla_{\alpha} h_{\nu)}^{\ \alpha} 
 & + \frac{1}{3} \nabla_{\alpha}\nabla_{\beta} h^{\alpha\beta} g_{\mu\nu} 
+ \frac{1}{3} ( \nabla_{\mu}\nabla_{\nu}
-   g_{\mu\nu} \Box) \,h
 \\ & 
+ R_{\mu\alpha\beta\nu} h^{\alpha\beta} - R_{\alpha(\mu} h_{\nu)}^{\alpha} 
+ \lambda \, C_{\mu\alpha\beta\nu} h^{\alpha\beta} =0.
\end{split}
\end{equation*}
where $h=h_{\mu}^\mu$ is the field trace, $\lambda$ is a free parameter and $C_{\mu\alpha\beta\nu}$ is the Weyl tensor. 
This equation is gauge dependent, which indicates the presence of both spin-0 and spin-2 components of $h_{\mu\nu}$ \cite{Barut:1982nj}.  In other words, the graviton field can be made conformally invariant - then propagates on light-cones - by adding a scalar component.
Moreover and for the sake of consistency let $h_{\mu\nu}=\phi g_{\mu\nu}$, where $\phi$ is a scalar field of weight $-1$, then the above equation leads as expected to the conformally invariant scalar field equation $(\Box +\frac{R}{6}) \phi =0$. 

%Note that  the 
%This new equation reduces to that given by Drew and Gegenberg in Minkowski space \cite{Drew:1980yk}.
%When the field is traceless the above equation reduces to the Deser and Nepomechie equation in constant curvature spaces\cite{Deser:1983tm}. 
%Moreover, if the Lorenz gauge is applied, $\nabla_{\mu}h_{\ \nu}^\mu=0$, the above equation reduces to that given by Grishchuk and Yudin \cite{Grishchuk:1980hb}  and is conformally invariant only for a restricted set of Weyl rescalings. This is because the conformal invariance of the Lorenz gauge requires $h^{\mu\nu}\partial_{\mu} \Omega=0$. This was the main argument against this result, led by Bekenstein and Meisels \cite{Bekenstein:1980jw}. However, this equation does not suffer from such a restriction and is perfectly conformally invariant.

This new equation extends and fixes the limitations of the Grishchuk and Yudin equation since it does not require the Lorenz gauge, $\nabla_{\mu}h_{\ \nu}^\mu=0$. Indeed, the latter constraint is conformally invariant for a restricted set of Weyl rescalings verifying $h^{\mu\nu}\partial_{\mu} \Omega=0$. This was the main argument against Grishchuk-Yudin's equation, led by Bekenstein and Meisels \cite{Bekenstein:1980jw}.
Note also that the above equation reduces to Drew-Gegenberg's equation in Minkowski space and, when the field is traceless, to Deser-Nepomechie's equation in constant curvature spaces.

This article is organized as follows. 
The next section underlines some features of Weyl geometry and states the Weyl-to-Riemann method.
Sec. \ref{application} is dedicated to applications. 
Sec. \ref{scalar} expands first, second and fourth order conformally invariant scalar field equations. 
Sec.  \ref{vector} deals with vector fields. In particular the Eastwood-Singer gauge is recovered.
In Sec. \ref{spin-2} the method is applied to construct the most general conformally invariant equation for a spin-2 field. %The third-rank tensor (or Fierz) representation is also explored.
Conclusions and outlooks are presented in Sec. \ref{conclusion}.
Finally, Appdx. \ref{weyl-tensors}  derives the Weylian geometrical tensors and Appdx. \ref{formulas}
provides some useful formulas.

We work in $d=4$ dimensions but the Weyl-to-Riemann method is, in principle, valid for any dimension $d\geq2$. The spaces are torsion free and equipped with metrics of signature ($+,---$). 
We use a tilde $(\sim)$ to indicate objects defined in a Weyl space and distinguish them from their Riemannian counterparts.

\newpage
\section{From Weyl to Riemann}\label{}
\subsection{Some features of Weyl geometry}\label{weyl-geometry} %geometrical framework
In 1918, Herman Weyl invented a new geometry in order to unify gravitation and electromagnetic forces. 
This goal was not reached but Weyl had exhibited a simple generalization of Riemann geometry and had by the way introduced gauge trasformations. These turned out to be fundamental concepts for both physics and mathematics \cite{Kastrup}.

A Weyl space is a real manifold equipped with two independent and equally fundamental objects:  the metric tensor $g$ and the Weyl vector  $W$.
%, when Riemann geometry is completely characterized by $g$.
%\begin{eqnarray*}
%Riemann \ geometry: & g_{\mu\nu},
%\\
%Weyl \ geometry: & [g_{\mu\nu}, W_{\mu}].
%\end{eqnarray*}
%The notation $[g,W]$ means that a Weyl space is a conformal equivalent class.
%In the framework of Weyl geometry, c
In Weyl geometry, conformal transformations mean gauge-Weyl transformations which transform both $g$ and $W$,
\begin{eqnarray}
g_{\mu\nu} &\to&  \bar{g}_{\mu\nu}  =  \Omega^{2} \ g_{\mu\nu},  \label{weyl}
\\
W_{\mu} &\to& \bar{W}_{\mu} = W_{\mu} + \Omega_{\mu}, \label{gauge-weyl}
\end{eqnarray}
where $\Omega$ is a real and smooth function and $\Omega_{\mu}=  \partial_{\mu} ln \Omega$.
The transformation (\ref{weyl}) alone is called rigid-Weyl (or Weyl) rescaling and defines conformal transformations in Riemann geometry.

%\subsection{Weyl connection}
There is a unique symmetric, torsion free and linear connection one can construct in Weyl geometry  \cite{Folland}. It is the Weyl connection,
\begin{equation}\label{weyl-connection}
\tilde \Gamma_{\mu\nu}^{\lambda} = \Gamma_{\mu\nu}^{\lambda} - W_{\mu\nu}^{\lambda} , \quad with
\end{equation}
\begin{eqnarray}
\Gamma_{\mu\nu}^{\lambda} &=& \frac{1}{2} g^{\lambda\tau}(g_{\tau(\mu,\nu)}  - g_{\mu\nu,\tau} ),\label{christoffel}
\\
W_{\mu\nu}^{\lambda}  &=& \delta^{\lambda}_{(\mu} W_{\nu)}  - g_{\mu\nu}W^{\lambda},\label{connection-W}
\end{eqnarray}
where $ W^{\lambda}= g^{\lambda\tau}W_{\tau}.$
As a consequence,  the Weyl connection is conformally (\ref{weyl}+\ref{gauge-weyl}) invariant,
\begin{equation}\label{inv-connection}
\tilde \Gamma \rightarrow \bar{\tilde \Gamma} = \tilde \Gamma.
\end{equation}
This is why conformally related spaces represent in fact the same Weyl space, which is a conformal equivalence class $[g,W]$. 
Another important consequence of the definition (\ref{weyl-connection}) is that  the associated covariant derivative is not metric compatible,
\begin{equation}\label{non-compatible}
\tilde\nabla_{\mu} \, g_{\alpha\beta} = \nabla_{\mu} \, g_{\alpha\beta} + W_{\mu(\alpha}^\lambda g_{\beta)\lambda} =
2 W_{\mu} \, g_{\alpha\beta} \neq 0,
\end{equation}
where $\nabla_{\mu}$ is defined by the Christoffel symbol (\ref{christoffel}).
This means that lengths are not preserved by parallel transport. In Weyl geometry, only angles (ratios of lengths) are meaningful and lengths can only be compared locally. The Weyl geometry is a "truly" differential geometry and this is in fact what H. Weyl was looking for \cite{ORaifeartaigh:2000}.

%\subsection{Weyl covariant derivative}\label{weyl-derivative}
In place of  $\tilde \nabla_{\mu}$, one can use the Weyl covariant derivative defined through
\begin{equation}\label{weyl-derivative}
\begin{array}{lll}
D_{\mu} T^{\alpha...}_{\beta...} 
& =&   (\tilde \nabla_{\mu} - w \, W_{\mu}) \  T^{\alpha...}_{\beta...}
%\\
%&=&  
=( \nabla_{\mu} - w \, W_{\mu}) \ T^{\alpha...}_{\beta...}   + \  W^{\lambda}_{\mu\beta} \ T^{\alpha...}_{\lambda...} \ + ...
 - \ W^{\alpha}_{\mu\lambda} \ T^{\lambda...}_{\beta...} - ...,
\end{array}
\end{equation}
where $T$ is a given conformal tensor field of rank $(r,s)$ and conformal weight $w=w(T)$.
Under conformal transformations, conformal tensors are mapped like
\begin{equation}\label{tensor}
T \to \Omega^{w}\, T.
\end{equation}
The Weyl covariant derivative is metric compatible,
$D_{\mu} \, g_{\alpha\beta} = 0,$
and is doubly covariant, according to conformal transformations (thanks to to the invariance of the Weyl connection) and diffeomorphisms.
Moreover it is linear and verifies the Leibniz rule.

\subsection{The Weyl-to-Riemann method}\label{method}
The purpose of the Weyl-to-Riemann method is to provide a simple and systematic algorithm which translates the problem of constructing conformally invariant equations in arbitrary Riemann spaces into a set of algebraic equations. 
The method is based on the two following features of Weyl geometry:

\begin{itemize}
\item 
Using the Weyl covariant derivative (\ref{weyl-derivative}) and the Weyl geometrical tensors given in Appdx. \ref{weyl-tensors}, all homogeneous equations written in a Weyl space are conformally invariant. Homogeneous here means that the different terms have the same conformal weight.
\item
When the Weyl vector is gradient,
$%\begin{equation*}
W_{\mu} = \partial_{\mu} \ln K,
$ %\end{equation*}
where $K$ is a real and smooth function, the Weyl space is then called Weyl Integrable Space (or WIS) and the Weyl connection (\ref{weyl-connection}) reduces to
$ %\begin{equation*}
\tilde \Gamma^{\lambda}_{\mu\nu} = \bar \Gamma^{\lambda}_{\mu\nu},
$ %\end{equation*}
where $ \bar \Gamma^{\lambda}_{\mu\nu}$ is the Christoffel connection of the metric tensor $\bar{g}_{\mu\nu} = K^{-2}g_{\mu\nu}$.
The covariant derivative becomes 
%$\bar \nabla$ associated to $\bar \Gamma$ verifies
%$\bar \nabla_{\mu} \, \bar{g}_{\alpha\beta} = 0$.
 metric compatible, which means that the WIS $[g_{\mu\nu}, \partial_{\mu} \ln K]$ is in fact a Riemann space equipped with the metric $\bar{g}_{\mu\nu}$. 
Note that some authors use WISs in a different way \cite{Hochberg:1990xp, Novello:2009}.
\end{itemize}

In practice, to construct a Riemannian conformally invariant n-order equation for a tensor field $T$ of weight $w$, one starts by writing the most general homogeneous equation in a Weyl space, say
\begin{equation}\label{emethod}
 a_{1} \tilde D_{1} + a_{2} \tilde D_{2} + \dots + b_{1} \tilde U_{1} + b_{2} \tilde U_{2}  + \dots = 0,
\end{equation}
where $\tilde D_{i}$ are derivative parts acting on $T$, $\tilde U_{i}$ are geometrical parts and $a_{i}, b_{i}$ are free parameters.
As stated above, this equation is naturally conformally (gauge-Weyl) invariant in a Weyl space. 
The game consists in looking for combinations $\{w, a_{i}, b_{i}\}$ in order to obtain, when reducing to a Riemann space, by imposing $W_{\mu} = \partial_{\mu} \ln K$, an  equation 
 \begin{equation}\label{emethodr}
a_{1}  D_{1} + a_{2}  D_{2} + \dots + b_{1}  U_{1} + b_{2}  U_{2}  + \dots = 0,
\end{equation}
 which is conformally (rigid-Weyl) invariant.
Several situations may occur:
\begin{enumerate} 
\item \label{cas1}
There is a combination $\{w, a_{i}, b_{i}\}$ which cancels the $W$ dependence of the equation (\ref{emethod}).
The latter is thus rigid-Weyl invariant in a Weyl space. Imposing $W$  to be gradient becomes ``invisible'' and the resulting Riemannian equation (\ref{emethodr}) still rigid-Weyl invariant, \textit{i.e.} conformally invariant.
\item
There is no combination $\{w, a_{i}, b_{i}\}$ which cancels the $W$ dependence of the equation (\ref{emethod}) and then three cases are possible:
\begin{enumerate}
\item\label{cas2a}
$W$ decouples when it has a gradient form. The equation (\ref{emethodr}) is then conformally invariant.
\item\label{cas2b}
An extra conformally invariant constraint cancels $W$ in (\ref{emethod}). The equation (\ref{emethodr}), together with the involved constraint, becomes conformally invariant.
\item \label{cas2c}
There is no way to get rid of $W$. The conformal invariance is broken when  going to a Riemann space.
\end{enumerate}
\end{enumerate}

\section{Applications}\label{application}
The Weyl-to-Riemann method is applied to construct conformally invariant equations for some integer-spin conformal tensor fields $T$.  In Riemann geometry, these fields have canonical weights which are related to the spin $s$  by the relation 
\begin{equation}\label{weight-spin}
w=s-1.
\end{equation}

Otherwise stated, we are not considering derivatives of the Weyl geometrical tensors.

\subsection{Conformally invariant scalar field equations}\label{scalar}
\subsubsection{First order equation}
The conformally invariant first order equation acting on a  scalar field $\phi$ defined in a Weyl space reads
\begin{equation}
D_{\mu} \, \phi = (\nabla_{\mu} - w W_{\nu} ) \, \phi = 0.
\end{equation}
The Weyl vector decouples when $w=0$ and the resulting Riemannian equation 
\begin{equation}
\nabla_{\mu} \phi=0
\end{equation}
 is then conformally invariant, but not with the canonical weight (\ref{weight-spin}). This corresponds to the situation \ref{method}.\ref{cas1}.

\subsubsection{Second order equation}
The most general second order scalar field equation in a Weyl space involves the second-rank geometrical tensors, it reads
\begin{equation}\label{}
g^{\mu\nu}(D_{\mu} D_{\nu}  + b_{1} \tilde{R}_{\mu\nu} + b_{2} \tilde R g_{\mu\nu} ) \,  \phi  = 
(D^2  +\alpha \tilde{R} ) \,  \phi = 0,
\end{equation}
where $\alpha=b_{1}+4b_{2}$.
This equation is expanded as
\begin{equation}\label{scalar-2}
\Big( \Box + \alpha R -2(1+w) W.\nabla -(6\alpha+w) \nabla.W +  (w^2+2w+6\alpha) W^2 \Big) \, \phi
= 0,
\end{equation}
where $\Box = g^{\mu\nu}\nabla_{\mu}\nabla_{\nu}$ and all other contractions are performed using $g_{\mu\nu}$. 
We can eliminate $W$ from this equation iff 
\begin{equation}
\begin{split}
1+w & = 0,
\\
6\alpha +w & =0,
\\
w^2+2w+6\alpha &=0.
\end{split}
\end{equation}
This system has a unique solution 
\begin{equation}
w=-1, \qquad \alpha=\frac{1}{6}.
\end{equation}
We are here again in the situation \ref{method}.\ref{cas1}
 The equation (\ref{scalar-2}) then reduces to the usual conformally invariant scalar field equation in Riemann spaces,
\begin{equation}\label{scalar-field-equation}
(\Box + \frac{1}{6} R) \phi = 0.
\end{equation}

%Remark:
%Let us fix $w=-1$, otherwise the term $W.\nabla$ remains in the equation, and relax the constraint on $\alpha$. Setting  $\alpha = 1/6  + \theta/6$ with $\theta \neq 0$, the equation (\ref{scalar-2}) reduces to
%$
%\left( \Box +\frac{1}{6} (R+\theta \tilde R) \right) \, \phi = 0.
%$
%The Weyl vector (present in $\tilde R$) does not cancel even when it is a gradient. We are in the situation \ref{method}.\ref{cas2c}.
%We do not obtain a conformally invariant equation when going into a Riemann space.
%Instead, we can set $6m^2  =  \theta \tilde R = \theta [ R +  6(W^2 - \nabla.W) ]$ and yields to the equation
%\begin{equation}
%\left( \Box +\frac{1}{6} R  + m^2 \right) \, \phi = 0,
%\end{equation}
%where $m$ can be interpreted as a mass term breaking the conformal invariance. In this case, the field $\phi$ reduces to a  massive scalar field in a Riemannian space. 

\subsubsection{Fourth order equation}
Let us now consider a fourth order conformally invariant scalar equation in a Weyl space given by
\begin{equation}\label{scalar-4}
 D_{\mu} \Big( D^{\mu}D^{\nu}  + a \tilde{R}^{\mu\nu} + b \tilde{R} g^{\mu\nu} \Big) D_{\nu} \,  \phi =0.
\end{equation}
Note that this equation involves derivatives of the geometrical tensors.
A straightforward calculation  shows that there is a unique combination,
\begin{equation}
w=0, \qquad a=-2, \qquad b=2/3,
\end{equation}
which cancels the Weyl vector provided it verifies the free Maxwell equation 
\begin{equation}\label{max-w}
\nabla_{\mu}\nabla^{[\mu}W^{\nu]}=0.
\end{equation}
The equation (\ref{scalar-4}) thus reduces to 
\begin{equation}\label{4-order-scalar}
 \nabla_{\mu} \Big( \nabla^{\mu}\nabla^{\nu}  + {S}^{\mu\nu} \Big) \nabla_{\nu} \,  \phi =0,
\end{equation}
where the Eastwood-Singer tensor reads
\begin{equation}\label{ES-tensor}
S^{\mu\nu} = -2 R^{\mu\nu} + \frac{2}{3} R g^{\mu\nu}.
\end{equation}
When $W_{\mu}$ is a gradient, \textit{i.e.} a pure gauge verifying identically (\ref{max-w}), the equation (\ref{4-order-scalar}) becomes conformally invariant in a Riemann space with a vanishing conformal weight.
Though in a different way, this equation was derived by Eastwood and Singer in \cite{Eastwood}.

\subsection{Conformally invariant vector field equations}\label{vector}
Let us apply the Weyl-to-Riemann method to spin-1 vector field. We first recover the conformal invariance of the Riemannian Maxwell equation and then look for some conformally invariant gauge fixing conditions.

\subsubsection{Maxwell equation}
The Maxwell equation in a Weyl space for a one-form $A_{\mu}$ reads 
\begin{equation}
 D^{\mu} D_{[\mu}A_{\nu]} = 0.
 \end{equation}
It is conformally invariant for any  weight $w=w(A)$ and extends into
\begin{equation}
 \nabla^{\mu} F_{\mu\nu} - w \Big( W^{\mu}F_{\mu\nu}  + (\nabla^{\mu} - w W^{\mu}) (W_{\mu}A_{\nu} - W_{\nu}A_{\mu})  \Big) =0.
\end{equation}
where $F_{\mu\nu} = \nabla_{[\mu} A_{\nu]}$.
For $w=0$, the Weyl vector cancels. Thus one obtain the conformally invariant Maxwell equation,  $\nabla^{\mu}F_{\mu\nu}=0$, with the usual vanishing conformal weight in a Riemann space.

\subsubsection{Lorenz gauge}\label{lorenz-gauge}
The Weylian Lorenz gauge, $ D.A= g^{\mu\nu} D_{\mu}A_{\nu}=0$, yields
\begin{equation}
\nabla.A - (2+w) W.A = 0
\end{equation}
The Weyl vector decouples for a non canonical weight, $w=-2$ or when the constraint $W.A =0$ is applied. The latter is conformally invariant for a restricted set of Weyl rescalings verifying
\begin{equation}\label{perp-omega}
\Omega.A =0.
\end{equation}
Accordingly the Riemannian Lorenz gauge, $\nabla.A=0$, is not completely conformally invariant for a canonical weight, $w=0$.

\subsubsection{Eastwood-Singer gauge}\label{confo-gauge}
The simplest  third order differential equation for the field $A_{\mu}$ of a vanishing weight in a Weyl space reads
\begin{equation}
 D^2 D.A  =  0.
\end{equation}
This equation contains only derivative terms but is perfectly conformally invariant. It expands as
\begin{equation}\label{only-derivatives}
\begin{split}
D^{2} D.A &= \nabla^{\mu} (\nabla_{\mu} + 2 W_{\mu}) \, (\nabla^{\nu} - 2 W^{\nu})\, A_{\nu}
\\ & = \nabla_{\mu} (\nabla^{\mu}\nabla^{\nu} + S^{\mu\nu}) A_{\nu} - 2 W^{\nu} \nabla^{\mu} F_{\mu\nu}
 \\ & = 0,
 \end{split}
\end{equation}
where  $S^{\mu\nu}$ (\ref{ES-tensor}) is the Eastwwod-Singer  tensor.
We have an interesting situation corresponding to the case \ref{method}.\ref{cas2b}. Indeed, the equation (\ref{only-derivatives}) contains $W$ but when $A_{\mu}$ verifies the Maxwell equation the Weyl vector decouples. The equation  (\ref{only-derivatives}) then reduces to the Eastwood-Singer gauge 
\begin{equation}\label{ES-gauge}
\nabla_{\mu} (\nabla^{\mu}\nabla^{\nu} + S^{\mu\nu}) A_{\nu} = 0,
\end{equation}
which is a conformally invariant but only  when acting on the Maxwell equation solutions.
This gauge was used for instance in \cite{Esposito:2008dw, pconf3, pconf5}.
Note that for a pure gauge field $A_{\nu}=\nabla_{\nu}\phi$, the equation (\ref{ES-gauge}) reduces to the equation (\ref{4-order-scalar}).

\subsection{Conformally invariant spin-2 field equation}\label{spin-2}
Let us turn to the spin-2 field which describes weak gravitational fields and in particular gravitational waves in a given curved background.
It is well known that the linearized Einstein equations are not conformally invariant (which is also true for the full Einstein equations). 
It is also known that conformal invariance can be reached if gauge freedom is given up \cite{Deser:1983}.
We show how the Weyl-to-Riemann method can be applied to construct conformally invariant equations for a spin-2 field in arbitrary Riemann spaces.

%\subsubsection{Spin-2 field in the second-rank tensor representation}\label{}
A spin-2 field is usually represented by a symmetric tensor field $h_{\mu\nu}=h_{\nu\mu}$. The trace and divergence are free. 
In order to exploit the entire capacity of the Weyl-to-Riemann method and increase the chances to find new results, let us write the most general second order spin-2 field equation in a Weyl space,
\begin{equation}\label{spin2}
\tilde D_{\mu\nu}  + \tilde  U_{\mu\nu} = 0,
\end{equation}
where the derivative part is given by
\begin{equation}\label{derivaticepart}
\tilde D_{\mu\nu}  = D^2 h_{\mu\nu} + a_1\, D_{(\mu} D^{\alpha} h_{\nu)\alpha} + a_2\, D_{\alpha}D_{\beta} h^{\alpha\beta} g_{\mu\nu}
+ a_3\, D_{\mu} D_{\nu} h + a_4\, D^2 h g_{\mu\nu},
\end{equation}
and the geometrical part reads
\begin{equation}\label{geopart}
\tilde U_{\mu\nu} = b_1\, \tilde R_{\mu\alpha\beta\nu} h^{\alpha\beta} + b_2\, \tilde R_{\alpha\beta} h^{\alpha\beta}g_{\mu\nu} 
+  b_3\, \tilde R_{\alpha(\mu} h_{\nu)}^{\alpha} +  b_4\, \tilde R\, h_{\mu\nu} +  b_5\, \tilde R_{\mu\nu} h +  b_6\, \tilde R \, g_{\mu\nu}h.
\end{equation}
The free parameters $\{a_i, b_i\}$ are real constants.
The different terms have the same conformal weight $w-2$, which ensures conformal homogeneity. 
The appearance of the non conformally invariant tensor $\tilde R$ in (\ref{geopart}) comes from the contraction of the conformally invariant tensor $\tilde R g_{\mu\nu}$.
Note that the covariant derivatives do not commute, thus using $D_{\alpha} D_{(\mu} g_{\nu)\beta}$ instead of $D_{(\mu} D_{\alpha} g_{\nu)\beta}$ slightly modifies the final form of the equation. However, this modification amounts to select a different set of the geometrical tensors parameters by means of  the identity (\ref{commutation-riemann}).
In the following we expand the equation (\ref{spin2}) then we look for combinations of the different parameters which cancels the Weyl vector dependence.

%\newpage	
%\subsubsection{Second solution}
%Let us relax the Lorenz gauge but keep the traceless condition (\ref{h=0}). Let us also impose the constraint $d=1/6$ of the previous subsection. Here again, we are looking for combination allowing to get rid of the Weyl vector in the equation (\ref{spin2}).

Considering first the derivative part (\ref{derivaticepart}) and trying to cancel the terms containing both $W$ and derivatives of $h_{\mu\nu}$, because these terms cannot be compensated by the geometrical part. We get the following
\begin{equation}
\begin{split}
\tilde D_{\mu\nu}  = D_{\mu\nu} \ & + 2(w-1) W^\alpha \nabla_{\alpha} h_{\mu\nu}
\\  &
			   + (2+a_1(w+2)) W^\alpha \nabla_{(\mu} h_{\nu)\alpha}
\\ &
			   - (2-a_1(w-4))  W_{(\mu} \nabla^\alpha h_{\nu)\alpha}
\\  &
			   +2(a_1+a_2(w+1)) W^{\alpha} \nabla^\beta h_{\alpha\beta} g_{\mu\nu}
\\ &
+ (a_1  + a_3(3-w)) \ h_{,(\mu} W_{\nu)}
\\&
+ (a_2-a_3 + 2a_4(1-w)) \ h_{,\alpha}W^\alpha g_{\mu\nu}
\\  &
			+ non \ derivative\ terms,
\end{split}
\end{equation}
The resulting algebraic system reads
\begin{equation*}
\begin{split}
w-1 &=0,
\\
2+e(w+2) &=0 ,
\\
2 - a_1(w-4)  &=0, 
\\
a_1+a_2(w+1) &=0,
\\
a_1  + a_3(3-w) &=0,
\\
a_2-a_3 + 2a_4(1-w) &=0.
\end{split}
\end{equation*}
This system has an infinite set of solutions, which can be put under the form
\begin{equation}
w=1, \qquad a_1=-\frac{2}{3} , \qquad a_2=\frac{1}{3}, \qquad a_3=\frac{1}{3}, \qquad a_4= \tau,
\end{equation}
where $\tau$ is a free parameter.
The resulting expression for the derivative part reads
\begin{equation}
\begin{split}
 \tilde  D_{\mu\nu}   =    D_{\mu\nu}   
 & + 2 \left( W_{\alpha} h^\alpha_{(\mu} W_{\nu)} +  h^\alpha_{(\mu} W_{\alpha;\nu)}  \right) 
 - ( W_{\alpha;\beta} + W_{\alpha} W_{\beta}) h^{\alpha\beta} g_{\mu\nu} +  (\nabla.W - 3 W^2) h_{\mu\nu} 
\\&  
+(\tau-\frac{1}{3}) (\nabla.W-W^2) h\, g_{\mu\nu} - h \Big( \frac{1}{2} W_{(\mu;\nu)} + W_{\mu}W_{\nu} - \frac{2}{3} g_{\mu\nu} W^2  \Big).
\end{split}
\end{equation}
The Weyl vector does not decouple completely. 
Also we add the geometrical part (\ref{geopart}) to get rid of it. The general equation (\ref{spin2}) expands into
\begin{equation}\label{general}
\begin{split}
 \tilde D_{\mu\nu}  + \tilde U_{\mu\nu}  
  =   D_{\mu\nu}  +  U_{\mu\nu} 
 & + (1 -2b_3 - 6b_4 ) \, \nabla.W h_{\mu\nu}
 \\& +  (-3 + b_1 +4b_3 + 6b_4 ) \, W^2 h_{\mu\nu}
\\& + (2-b_1-2b_3) \ \Big( W_{\alpha}W_{(\mu}  + W_{\alpha;(\mu} \Big) h_{\nu)}^{\ \alpha} 
\\&+  (b_1+2b_3)   \Big(  W_{\alpha;(\mu} - W_{(\mu;\alpha} \Big) h^\alpha_{\ \nu)}
\\
& + (-1+b_1-2b_2) \,  \left( W_{\alpha;\beta} + W_{\alpha}W_{\beta} \right) h^{\alpha\beta} g_{\mu\nu}  
\\
& + (-1+b_1-2b_5) \, \left( 1/2 W_{(\mu;\nu)} + W_{\mu}W_{\nu} \right) \, h
\\ 
& +  \left(1/3 - b_2 - b_5 + \tau + 6b_6 \right) \ \nabla.W  h\, g_{\mu\nu}
\\&+ \left(  2/3 - b_1 + 2 b_2+ 2b_5 - \tau +6b_6  \right) W^2  h\, g_{\mu\nu}
\\ & =0.
\end{split}
\end{equation}
Except the fourth line which cancels when $W$ is a gradient, eliminating $W$ leads to the following algebraic system
\begin{equation*}
\begin{split}
 1-2b_3 -6b_4 & =0 ,
\\
 -3 +b_1+4b_3+6b_4 & =0 ,
\\
-1+b_1 - 2b_2 & =0 ,
\\
-1+b_1-2b_5 &=0,
\\
2-b_1-2b_3 &=0,
\\
1/3 - b_2 - b_5 + \tau - 6b_6 &=0,
\\
2/3 - b_1 + 2 b_2+ 2b_5 - \tau +6b_6 &=0.
\end{split}
\end{equation*}
This system has an infinite set of solutions which can take the form
\begin{equation*}
b_1=\lambda+1, \quad b_2= \frac{\lambda}{2}, \quad b_3=\lambda-1, \quad b_4=\frac{\lambda}{6}, \quad b_5= \frac{\lambda}{2} , \quad 6b_6=  \tau - \lambda + 1/3,
\end{equation*}
where $\lambda, \tau$ are free parameters.
The equation (\ref{general}) becomes
\begin{equation}\label{gen-eq}
 D_{\mu\nu}  +  U_{\mu\nu}  + 2 h^\alpha_{\ (\mu}\, \tilde W_{\nu)\alpha } =0,
\end{equation}
where $\tilde W_{\mu\nu}$ is given by (\ref{w-tensor}) and the purely Riemann parts read
\begin{eqnarray}
D_{\mu\nu}  &=& \Box h_{\mu\nu} \ -\frac{2}{3}\, \nabla_{(\mu} \nabla^{\alpha} h_{\nu)\alpha} + \frac{1}{3}\, \nabla_{\alpha}\nabla_{\beta} h^{\alpha\beta} g_{\mu\nu} + \frac{1}{3} \nabla_{\mu} \nabla_{\nu} h + \tau \Box h g_{\mu\nu},
\label{deriv-part}
\\
 U_{\mu\nu}  &=&   R_{\mu\alpha\beta\nu} h^{\alpha\beta} - R_{\alpha(\mu} h_{\nu)}^{\alpha} 
+ \frac{ 1+3\tau}{18} R \, g_{\mu\nu}h
+ \lambda \, C_{\mu\alpha\beta\nu} h^{\alpha\beta}.
\label{r-geopart}
\end{eqnarray}
The appearance of the Weyl tensor 
\begin{equation}
\begin{split}
C_{\mu\alpha\beta\nu}  =  R_{\mu\alpha\beta\nu} \, + \, \frac{1}{2} ( g_{\mu[\nu} R_{\alpha\beta]} + g_{\alpha[\beta} R_{\mu\nu]} )
- \frac{R}{6} (g_{\mu[\nu} g_{\alpha\beta]}),
\end{split}
\end{equation}
which is by itself conformally invariant with weight $+2$, reflects the presence of free parameters in the geometrical part.
The remaining $W$ dependence in the equation (\ref{gen-eq}) cancels when going to Riemann geometry.
Moreover, the contraction of  (\ref{gen-eq}) 
%leads to
%\begin{equation}
%(1+3\tau) (\Box +\frac{R}{6}) h =0,
%\end{equation}
 fixes one of the free parameter,
 \begin{equation}\label{tau}
 \tau=-\frac{1}{3}.
\end{equation}

We obtain the equation

\begin{equation}\label{sosso}
\begin{split}
 \Box \, h_{\mu\nu}  - \frac{2}{3} \nabla_{(\mu}\nabla_{\alpha} h_{\nu)}^{\ \alpha} 
 & + \frac{1}{3} \nabla_{\alpha}\nabla_{\beta} h^{\alpha\beta} g_{\mu\nu} 
+ \frac{1}{3} ( \nabla_{\mu}\nabla_{\nu}
-   g_{\mu\nu} \Box) \,h
 \\ & 
+ R_{\mu\alpha\beta\nu} h^{\alpha\beta} - R_{\alpha(\mu} h_{\nu)}^{\alpha} 
+ \lambda \, C_{\mu\alpha\beta\nu} h^{\alpha\beta} =0.
\end{split}
\end{equation}
This equation is conformally invariant in Riemann geometry, with weight $-1$, for the spin-2 field represented by the symmetric tensor $h_{\mu\nu}$ with the canonical weight $+1$.
When $h_{\mu\nu}=\phi g_{\mu\nu}$, where $\phi$ is conformal scalar field with the canonical weight $-1$,
this equation reduces to the equation (\ref{scalar-field-equation}).  This proves the consistency of our result.
Note that this equation is gauge variant, which indicates the presence of both spin-0 and spin-2 components of $h_{\mu\nu}$ \cite{Barut:1982nj}. In other words, the $h_{\mu\nu}$ field can be made conformally invariant - to propagate on light-cones - by adding a scalar component.

The above equation reduces to Drew-Gegenberg's equation in Minkowski space \cite{Drew:1980yk}
and to Deser-Nepomechie's equation in constant curvature spaces \cite{Deser:1983tm}. Actually, Deser and Nepomechie presented a conformally invariant equation in general curved spaces but this was extended from an equation valid in de Sitter space. This explains why their equation lacks some terms of (\ref{sosso}) for a traceless field.
 
When the field is traceless, $h=0$, and divergentless, $\nabla_{\mu}h_{\ \nu}^\mu=0,$  the equation (\ref{sosso}) takes the simple form
\begin{equation}\label{grishchuk}
 \Box \, h_{\mu\nu} 
+ R_{\mu\alpha\beta\nu} h^{\alpha\beta} - R_{\alpha(\mu} h_{\nu)}^{\alpha} 
+ \lambda \, C_{\mu\alpha\beta\nu} h^{\alpha\beta} =0.
\end{equation}
This equation was given by Grishchuk and Yudin  \cite{Grishchuk:1980hb} and has been afterward criticized by Bekenstein and Meisels \cite{Bekenstein:1980jw}. Indeed, contrary to the traceless constraint which is fully conformally invariant, the Lorenz gauge is invariant only for a restricted set of Weyl rescalings. This is because conformal invariance of $\nabla_{\mu}h_{\ \nu}^\mu=0,$ requires $h^{\mu\nu}\partial_{\mu} \Omega=0$. This is equivalent to the constraint (\ref{perp-omega}) necessary to make the Lorenz gauge conformally invariant in the case of Maxwell field. 
This was the main argument against the Grishchuk and Yudin equation in particular and conformal spin-2 fields in general. Fortunately, the equation (\ref{sosso}) does not suffer from such a restriction and is perfectly conformally invariant.

\section{Concluding remarks}\label{conclusion}
The Weyl-to-Riemann method  was presented.  This method allows to construct conformally invariant equations in arbitrary Riemann spaces. It exploits two features of Weyl geometry. First, all homogeneous equations defined in a Weyl space are conformally invariant. Second, a simple procedure reduces a Weyl space to a Riemann space. The method selects those equations whose conformal invariance is preserved when reducing to a Riemann space.

The purpose of this method is to provide a simple, systematic and practical procedure to construct conformally invariant equations in general Riemann spaces. The major part of this article gathered practical applications. Scalar, vector and spin-2 fields were considered in order to demonstrate the efficiency of the Weyl-to-Riemann method. These applications showed that contrary to the Ricci-gauging method our technique is more practical, is not limited to second-order differential equations and is more suitable to find new results.

In particular a new conformally invariant spin-2 field second-order equation (\ref{sosso}) was exhibited. This generalizes all other known equations and allows to avoid the usual critics against conformal spin-2 field.
The physical content of this equation remains to be explored and experimentally tested using for instance gravitational waves future observations.

The Weyl-to-Riemann method can be applied to half-integer spin fields, higher-order and higher-rank tensor field equations. An interesting application would be related to the Fierz (third-rank tensor) representation of spin-2 field. Indeed, the Fierz representation affords remarkable features in the non conformally invariant sector \cite{Novello:2002bp} and one would expect important results in the conformally invariant one.

%\modif{
%This equation is to be compared with the linearized Einstein's equation \cite{Sokoowski:2007fk}
%\begin{equation}\label{}
% \begin{split}
%& \Box h_{\mu\nu} 
% - \, \nabla_{(\mu} \nabla^{\alpha} h_{\nu)\alpha} 
%+ \nabla_{\alpha}  \nabla_{\beta} h^{\alpha\beta} g_{\mu\nu}
% +   \, (\nabla_{\mu} \nabla_{\nu} - g_{\mu\nu}\Box ) h
%\\
%&  + 2 R_{\alpha\mu\nu\beta} h^{\alpha\beta}  + ( h_{\mu\nu} g_{\alpha\beta} + h_{\alpha\beta} g_{\mu\nu}) R^{\alpha\beta}
%+ (R_{\mu\nu} - \frac{1}{2}R\, g_{\mu\nu})\, h
%= 0.
%\end{split}
%\end{equation}
%}

\appendix

\section{Weyl geometrical tensors}\label{weyl-tensors}
Curvature tensors completely determine the geometry in both Riemann and Weyl spaces. They respectively read \cite{Constantinescu:2009fk}
\begin{eqnarray}
R^{\lambda}_{\ \mu\nu\beta} & = \partial_{\nu} \Gamma^{\lambda}_{\mu\beta} -  \partial_{\beta} \Gamma^{\lambda}_{\mu\nu} +  \Gamma^{\tau}_{\mu\beta}  \, \Gamma^{\lambda}_{\nu\tau} -  \Gamma^{\tau}_{\mu\nu}  \, \Gamma^{\lambda}_{\beta\tau}, \label{R-riemann}
\\
\tilde{R}^{\lambda}_{\ \mu\nu\beta} & = \partial_{\nu} \tilde\Gamma^{\lambda}_{\mu\beta} -  \partial_{\beta} \tilde\Gamma^{\lambda}_{\mu\nu}
+ \tilde \Gamma^{\tau}_{\mu\beta}  \, \tilde\Gamma^{\lambda}_{\nu\tau}
-  \tilde\Gamma^{\tau}_{\mu\nu}  \, \tilde\Gamma^{\lambda}_{\beta\tau} . \label{ssss}
\end{eqnarray}
The Weylian curvature tensor is conformally invariant since does the Weyl connection (\ref{inv-connection}). It expands into
\begin{equation}\label{riemann}
\begin{split}
\tilde{R}^{\lambda}_{\ \mu\nu\beta}  & =  R^{\lambda}_{\ \mu\nu\beta}
  + \nabla_{\nu} W_{\mu\beta}^\lambda  - \nabla_{\beta} W_{\mu\nu}^\lambda
 + W_{\mu\beta}^\tau W_{\nu\tau}^\lambda -  W_{\mu\nu}^\tau W_{\beta\tau}^\lambda  ,
 \\
 & =  R^{\lambda}_{\ \mu\nu\beta} -  \Big( \delta^\lambda_{\ \beta} W_{\mu;\nu} + W^\lambda_{\ ;\beta} g_{\mu\nu}    
 - \delta^\lambda_{\ \nu} W_{\mu;\beta} -  W^\lambda_{\ ;\nu} g_{\mu\beta} 
 +  \delta^\lambda_{\ \mu} W_{[\beta;\nu]}
  \Big)
  \\ & 
  \qquad\qquad - \Big(   \delta^\lambda_{\ \beta} W_{\mu} W_{\nu}  +  W^\lambda W_{\ \beta} g_{\mu\nu} 
  -  \delta^\lambda_{\ \nu} W_{\mu} W_{\beta} -  W^\lambda W_{\ \nu} g_{\mu\beta} 
    + g_{\mu[\beta} \delta^\lambda_{\ \nu]} 
     \Big).
\end{split}
\end{equation}
%where $\nabla_{\mu}$ refers to the riemannian covariant derivative related to the connection $\Gamma_{\mu\nu}^\lambda$.
%
This tensor possesses fewer algebraic symmetries than its Riemannian counterpart. As  a consequence  the Weyl curvature tensor has two possible contractions. The first contraction is $\tilde{W}_{\nu\beta} = \frac{1}{4} \tilde{R}^{\lambda}_{\ \lambda\mu\nu}$, which reads
\begin{equation}\label{w-tensor}
\begin{split}
\tilde{W}_{\mu\nu}  & = 
\nabla_{\mu} W_{\nu}  - \nabla_{\nu} W_{\mu} 
 = \partial_{\mu} W_{\nu} - \partial_{\nu} W_{\mu}.
\end{split}
\end{equation}
%It is the Faraday tensor for the Weyl vector (see the remark at the end of the section \ref{lengths}). 
This tensor is  antisymmetric, conformally invariant and is independent from the metric tensor. 
Moreover it is trace free and vanishes identically in a Riemann space (when $W$ is gradient). Also it is not very useful for the Weyl-to-Riemann method.

The second contraction of (\ref{riemann}) leads to the Weylian Ricci tensor $ \tilde{R}_{\mu\nu}  =\tilde{R}^{\lambda}_{\ \mu\lambda\nu}$. It is not symmetric and can be safely redefined under the symmetric form $\tilde{R}_{\mu\nu} = \tilde{R}^{\lambda}_{\ \mu\lambda\nu} - W_{\mu\nu} 
$. The latter is related to its Riemannian counterparts as follows,
\begin{equation}\label{ricci-weyl}
\begin{array}{lll}
\tilde{R}_{\mu\nu}  = &  R_{\mu\nu} 
 - \left( \nabla W  - 2 W^2 \right) g_{\mu\nu}  
- \left( \nabla_{(\mu} W_{\nu)}+ 2 W_{\mu} W_{\nu} \right).
\end{array}
\end{equation}
%where $R_{\mu\nu}$ is the Riemannian Ricci tensor.
The contraction of $\tilde R_{\mu\nu}$ produces the Weyl Ricci scalar %with a metric tensor, say $g^{\mu\nu}$ yields $R=g^{\mu\nu}R_{\mu\nu}$, which takes the form
\begin{equation}\label{R}
\tilde{R}(g)  = g^{\mu\nu} \tilde{R}_{\mu\nu}  = R - 6 (\nabla W -  W^2).
\end{equation}
This tensor is not conformally invariant since it  it has a non vanishing conformal weight. Indeed, it transformas like 
$
\tilde{R} \to \Omega^{-2} \tilde{R}.
$
This is related to the fact that the Ricci scalar, unlike the Riemann and Ricci tensors, is not defined only by a connection but needs a metric, \textit{i.e.} a scale.
However, the tensor
$%\begin{equation}\label{rg}
\tilde{R}g_{\mu\nu}  $ % = [ R - 6 (\nabla W -  W^2) ] g_{\mu\nu},
%\end{equation}
is conformally invariant. %and will be extensively used in what follows. 

Thus, the relevant Weylian geometrical tensors for the Weyl-to-Riemann method are $\tilde R_{\mu\nu\alpha\beta}, \tilde R_{\mu\nu}$ and $\tilde R g_{\mu\nu}$.

\section{Some useful formulas}\label{formulas}
Let $h_{\mu\nu}$ be a symmetric second-rank tensor with trace $h=h_{\mu}^{\ \mu}$. Here are some useful formulas. 

\begin{equation}
\begin{split}
 D^2 h_{\mu\nu} =   \ \Box \,  h_{\mu\nu} 
 &
 -2(w-1) h_{\mu\nu;\alpha} W^\alpha
 + (w^2-2w-2) W^2 h_{\mu\nu}
 - (w-2) \nabla.W h_{\mu\nu}
 \\&
 - 2 W_{\alpha} h^\alpha_{\ (\mu;\nu)}
+ 2 W_{(\mu} h_{\nu)\alpha;}  \, ^{\alpha}
- 4  W_{(\mu} h_{\nu)\alpha;}  \, W^\alpha
+ 2 g_{\mu\nu} h_{\alpha\beta}W^\alpha W^\beta
\end{split}
\end{equation}

\begin{equation}
\begin{split}
D_{(\mu}D_{\alpha} h_{\nu)}^{\ \alpha} = 
& 
h^\alpha_{(\mu;\alpha\nu)}
-(w-4) W_{(\mu} h_{\nu)\alpha;}  \, ^{\alpha}
-2 g_{\mu\nu} h^{\alpha\beta} \, _{;\alpha\beta}
- (w+2) h^\alpha_{\ (\mu;\nu)} W_{\alpha}
\\&
+ (w^2-2w-8) h^\alpha_{\ (\mu} W_{\nu)} W_{\alpha}
-(w+2) w_{\alpha;(\mu} h_{\nu)}^{\ \alpha}
+ 2 (2+w) g_{\mu\nu} h_{\alpha\beta}W^\alpha W^\beta
\\&
+ h_{;(\mu}W_{\nu)}
+ 2 h\Big( \frac{1}{2}W_{(\mu;\nu)} - g_{\mu\nu}W^2 + (4-w)W_{\mu}W_{\nu}  \Big)
\end{split}
\end{equation}

\begin{equation}
\begin{split}
D_{\mu}D_{\nu} h^{\mu\nu} =  h^{\mu\nu}\, _{;\mu\nu} 
&
- 2(1+w)h^{\mu\nu}_{\ \ ;\mu} W_{\nu} 
 + (2+w)h^{\mu\nu} (w W_{\mu}W_{\nu} - W_{\mu;\nu})
 \\&
+ h_{;\mu}W^\mu
+ h  (w W^2 - \nabla.W)
\end{split}
\end{equation}

\begin{equation}
\begin{split}
D_{\mu}D_{\nu} h=  h_{;\mu\nu}  
&
+ (3-w)  h_{;(\mu}W_{\nu)} - g_{\mu\nu} h_{;\alpha}W^\alpha
 \\&
+ (2-w) h\Big( \frac{1}{2}W_{(\mu;\nu)} - g_{\mu\nu}W^2 + (4-w)W_{\mu}W_{\nu}  \Big)
\end{split}
\end{equation}

Here are some useful expressions involving the geometrical tensors
\begin{equation}
\begin{split}
 \tilde R_{\mu\alpha\beta\nu} h^{\alpha\beta}   =R_{\mu\alpha\beta\nu} h^{\alpha\beta} 
& 
+ (W_{\alpha}W_{\beta} + \nabla_{\alpha} W_{\beta}) h^{\alpha\beta} g_{\mu\nu}
+ W^2 h_{\mu\nu}  
- h_{\ (\mu}^\alpha  (W_{\nu);\alpha } + W_{\alpha} W_{\nu)})
\\&
+h \Big( \frac{1}{2}W_{(\mu;\nu)} - g_{\mu\nu}W^2 + W_{\mu}W_{\nu}  \Big)
\end{split}
\end{equation}

\begin{equation}
\begin{split}
\tilde R_{\alpha(\mu} h_{\nu)}^{\ \alpha} = R_{\alpha(\mu} h_{\nu)}^{\ \alpha} 
& - 2  h_{\ (\mu}^\alpha  (W_{\nu);\alpha } + W_{\alpha} W_{\nu)})
 - 2 (\nabla.W - 2 W^2) h_{\mu\nu}
\end{split}
\end{equation}

\begin{equation}
\begin{split}
\tilde R_{\alpha\beta}\, h^{\alpha\beta}= R_{\alpha\beta}\, h^{\alpha\beta} 
- (2W_{\alpha}W_{\beta} + W_{\alpha;\beta}) h^{\alpha\beta} 
 - (\nabla.W - 2 W^2) h
\end{split}
\end{equation}

%The Weyl tensor in Weyl geometry reads
%\begin{equation}
%\tilde C_{\mu\alpha\beta\nu}  = \tilde R_{\mu\alpha\beta\nu} + \frac{1}{2} ( g_{\mu[\nu} \tilde R_{\alpha\beta]} + g_{\alpha[\beta} \tilde R_{\mu\nu]} )
%- \frac{\tilde R}{6} g_{\mu[\nu} g_{\alpha\beta]}.
%\end{equation}
%
%\begin{equation}
%\begin{split}
%\tilde C_{\mu\alpha\beta\nu} h^{\alpha\beta} =
% &  C_{\mu\alpha\beta\nu} h^{\alpha\beta}  - W_{\alpha(\mu} h_{\nu)}^{\ \alpha},
%\end{split}
%\end{equation}
%where $W_{\mu\nu}$ is given by (\ref{w-tensor}).
%

In Riemann geometry, the covariant derivatives do not commute. As a consequence one has the following identity
\begin{equation}\label{commutation-riemann}
h_{\ (\mu;\nu);\alpha}^{\alpha} - h_{\ (\mu;\alpha;\nu)}^{\alpha}
=  
2 R_{\mu\alpha\beta\nu} h^{\alpha\beta} 
+ R_{\alpha(\mu}h_{\nu)}^{\ \alpha},
\end{equation}

\section*{Acknowledgments}
We would like to thank Prof. Mario Novello for illuminating comments and the CNPq for financial support.

\baselineskip=10pt
\bibliography{Biblio}

\begin{thebibliography}{29}
\expandafter\ifx\csname natexlab\endcsname\relax\def\natexlab#1{#1}\fi
\expandafter\ifx\csname bibnamefont\endcsname\relax
  \def\bibnamefont#1{#1}\fi
\expandafter\ifx\csname bibfnamefont\endcsname\relax
  \def\bibfnamefont#1{#1}\fi
\expandafter\ifx\csname citenamefont\endcsname\relax
  \def\citenamefont#1{#1}\fi
\expandafter\ifx\csname url\endcsname\relax
  \def\url#1{\texttt{#1}}\fi
\expandafter\ifx\csname urlprefix\endcsname\relax\def\urlprefix{URL }\fi
\providecommand{\bibinfo}[2]{#2}
\providecommand{\eprint}[2][]{\url{#2}}

\bibitem[{\citenamefont{Deser and Nepomechie}(1984)}]{Deser:1983}
\bibinfo{author}{\bibfnamefont{S.}~\bibnamefont{Deser}} \bibnamefont{and}
  \bibinfo{author}{\bibfnamefont{R.~I.} \bibnamefont{Nepomechie}},
  \bibinfo{journal}{Annals Phys.} \textbf{\bibinfo{volume}{154}},
  \bibinfo{pages}{396} (\bibinfo{year}{1984}).

\bibitem[{\citenamefont{Boulanger}(2005)}]{Boulanger:2004}
\bibinfo{author}{\bibfnamefont{N.}~\bibnamefont{Boulanger}},
  \bibinfo{journal}{J.Math.Phys.} \textbf{\bibinfo{volume}{46}},
  \bibinfo{pages}{053508} (\bibinfo{year}{2005}), \eprint{hep-th/0412314}.

\bibitem[{\citenamefont{Shaynkman et~al.}(2006)\citenamefont{Shaynkman,
  Tipunin, and Vasiliev}}]{Shaynkman:fk}
\bibinfo{author}{\bibfnamefont{O.}~\bibnamefont{Shaynkman}},
  \bibinfo{author}{\bibfnamefont{I.}~\bibnamefont{Tipunin}}, \bibnamefont{and}
  \bibinfo{author}{\bibfnamefont{M.}~\bibnamefont{Vasiliev}},
  \bibinfo{journal}{Rev.Math. Phys.} \textbf{\bibinfo{volume}{18}}
  (\bibinfo{year}{2006}), \eprint{hep-th/0401086}.

\bibitem[{\citenamefont{Vasiliev}(2007)}]{Vasiliev:2007fk}
\bibinfo{author}{\bibfnamefont{M.~A.} \bibnamefont{Vasiliev}}
  (\bibinfo{year}{2007}), \eprint{0707.1085},
  \urlprefix\url{http://arxiv.org/abs/0707.1085}.

\bibitem[{\citenamefont{Vasiliev}(2010)}]{Vasiliev2010176}
\bibinfo{author}{\bibfnamefont{M.}~\bibnamefont{Vasiliev}},
  \bibinfo{journal}{Nuclear Physics B} \textbf{\bibinfo{volume}{829}},
  \bibinfo{pages}{176 } (\bibinfo{year}{2010}), ISSN \bibinfo{issn}{0550-3213}.

\bibitem[{\citenamefont{Gover et~al.}(2009)\citenamefont{Gover, Shaukat, and
  Waldron}}]{Gover:2008-tractors}
\bibinfo{author}{\bibfnamefont{A.}~\bibnamefont{Gover}},
  \bibinfo{author}{\bibfnamefont{A.}~\bibnamefont{Shaukat}}, \bibnamefont{and}
  \bibinfo{author}{\bibfnamefont{A.}~\bibnamefont{Waldron}},
  \bibinfo{journal}{Nucl.Phys.} \textbf{\bibinfo{volume}{B812}},
  \bibinfo{pages}{424} (\bibinfo{year}{2009}), \eprint{0810.2867}.

\bibitem[{\citenamefont{Fefferman and Graham}(2007)}]{Fefferman:2007}
\bibinfo{author}{\bibfnamefont{C.}~\bibnamefont{Fefferman}} \bibnamefont{and}
  \bibinfo{author}{\bibfnamefont{C.~R.} \bibnamefont{Graham}}
  (\bibinfo{year}{2007}), \eprint{0710.0919},
  \urlprefix\url{http://arxiv.org/abs/0710.0919}.

\bibitem[{\citenamefont{Manvelyan et~al.}(2007)\citenamefont{Manvelyan,
  Mkrtchyan, and Mkrtchyan}}]{Manvelyan:2007}
\bibinfo{author}{\bibfnamefont{R.}~\bibnamefont{Manvelyan}},
  \bibinfo{author}{\bibfnamefont{K.}~\bibnamefont{Mkrtchyan}},
  \bibnamefont{and}
  \bibinfo{author}{\bibfnamefont{R.}~\bibnamefont{Mkrtchyan}},
  \bibinfo{journal}{Phys.Lett.} \textbf{\bibinfo{volume}{B657}},
  \bibinfo{pages}{112} (\bibinfo{year}{2007}), \eprint{0707.1737}.

\bibitem[{\citenamefont{Erdmenger}(1997)}]{Erdmenger:1997}
\bibinfo{author}{\bibfnamefont{J.}~\bibnamefont{Erdmenger}},
  \bibinfo{journal}{Class.Quant.Grav.} \textbf{\bibinfo{volume}{14}},
  \bibinfo{pages}{2061} (\bibinfo{year}{1997}), \eprint{hep-th/9704108}.

\bibitem[{\citenamefont{Iorio et~al.}(1997)\citenamefont{Iorio, O'Raifeartaigh,
  Sachs, and Wiesendanger}}]{Iorio:1996}
\bibinfo{author}{\bibfnamefont{A.}~\bibnamefont{Iorio}},
  \bibinfo{author}{\bibfnamefont{L.}~\bibnamefont{O'Raifeartaigh}},
  \bibinfo{author}{\bibfnamefont{I.}~\bibnamefont{Sachs}}, \bibnamefont{and}
  \bibinfo{author}{\bibfnamefont{C.}~\bibnamefont{Wiesendanger}},
  \bibinfo{journal}{Nucl.Phys.} \textbf{\bibinfo{volume}{B495}},
  \bibinfo{pages}{433} (\bibinfo{year}{1997}).

\bibitem[{\citenamefont{Zumino}(1970)}]{Zumino}
\bibinfo{author}{\bibfnamefont{B.}~\bibnamefont{Zumino}},
  \bibinfo{journal}{Lectures on elementary particles and quantum field theory,
  Brandeis University Summer Institute} \textbf{\bibinfo{volume}{2}}
  (\bibinfo{year}{1970}).

\bibitem[{\citenamefont{Faci}(2012)}]{pconf7}
\bibinfo{author}{\bibfnamefont{S.}~\bibnamefont{Faci}},
  \bibinfo{journal}{arXiv:1206.4362}  (\bibinfo{year}{2012}).

\bibitem[{\citenamefont{Drechsler and Tann}(1999)}]{Drechsler:1999fk}
\bibinfo{author}{\bibfnamefont{W.}~\bibnamefont{Drechsler}} \bibnamefont{and}
  \bibinfo{author}{\bibfnamefont{H.}~\bibnamefont{Tann}},
  \bibinfo{journal}{Found.Phys.} \textbf{\bibinfo{volume}{29}},
  \bibinfo{pages}{1023} (\bibinfo{year}{1999}), \eprint{gr-qc/9802044}.

\bibitem[{\citenamefont{Barut and Xu}(1982)}]{Barut:1982nj}
\bibinfo{author}{\bibfnamefont{A.}~\bibnamefont{Barut}} \bibnamefont{and}
  \bibinfo{author}{\bibfnamefont{B.-W.} \bibnamefont{Xu}},
  \bibinfo{journal}{J.Phys.} \textbf{\bibinfo{volume}{A15}},
  \bibinfo{pages}{L207} (\bibinfo{year}{1982}).

\bibitem[{\citenamefont{Bekenstein and Meisels}(1980)}]{Bekenstein:1980jw}
\bibinfo{author}{\bibfnamefont{J.~D.} \bibnamefont{Bekenstein}}
  \bibnamefont{and} \bibinfo{author}{\bibfnamefont{A.}~\bibnamefont{Meisels}},
  \bibinfo{journal}{Phys.Rev.} \textbf{\bibinfo{volume}{D22}},
  \bibinfo{pages}{1313} (\bibinfo{year}{1980}).

\bibitem[{\citenamefont{Kastrup}(2008)}]{Kastrup}
\bibinfo{author}{\bibfnamefont{H.}~\bibnamefont{Kastrup}},
  \bibinfo{journal}{Annalen Phys.} \textbf{\bibinfo{volume}{17}},
  \bibinfo{pages}{631} (\bibinfo{year}{2008}), \eprint{0808.2730}.

\bibitem[{\citenamefont{Folland}(1970)}]{Folland}
\bibinfo{author}{\bibfnamefont{G.~B.} \bibnamefont{Folland}},
  \bibinfo{journal}{Journal of differential geometry}
  \textbf{\bibinfo{volume}{4}} (\bibinfo{year}{1970}).

\bibitem[{\citenamefont{O'Raifeartaigh and
  Straumann}(2000)}]{ORaifeartaigh:2000}
\bibinfo{author}{\bibfnamefont{L.}~\bibnamefont{O'Raifeartaigh}}
  \bibnamefont{and}
  \bibinfo{author}{\bibfnamefont{N.}~\bibnamefont{Straumann}},
  \bibinfo{journal}{Rev.Mod.Phys.} \textbf{\bibinfo{volume}{72}},
  \bibinfo{pages}{1} (\bibinfo{year}{2000}).

\bibitem[{\citenamefont{Hochberg and Plunien}(1991)}]{Hochberg:1990xp}
\bibinfo{author}{\bibfnamefont{D.}~\bibnamefont{Hochberg}} \bibnamefont{and}
  \bibinfo{author}{\bibfnamefont{G.}~\bibnamefont{Plunien}},
  \bibinfo{journal}{Phys.Rev.} \textbf{\bibinfo{volume}{D43}},
  \bibinfo{pages}{3358} (\bibinfo{year}{1991}).

\bibitem[{\citenamefont{Novello et~al.}(2011)\citenamefont{Novello, Salim, and
  Falciano}}]{Novello:2009}
\bibinfo{author}{\bibfnamefont{M.}~\bibnamefont{Novello}},
  \bibinfo{author}{\bibfnamefont{J.}~\bibnamefont{Salim}}, \bibnamefont{and}
  \bibinfo{author}{\bibfnamefont{F.}~\bibnamefont{Falciano}},
  \bibinfo{journal}{Int.J.Geom.Meth.Mod.Phys.} \textbf{\bibinfo{volume}{8}},
  \bibinfo{pages}{87} (\bibinfo{year}{2011}), \eprint{0901.3741}.

\bibitem[{\citenamefont{Eastwood and Singer}(1985)}]{Eastwood}
\bibinfo{author}{\bibfnamefont{M.~G.} \bibnamefont{Eastwood}} \bibnamefont{and}
  \bibinfo{author}{\bibfnamefont{M.}~\bibnamefont{Singer}},
  \bibinfo{journal}{Phys.Lett.} \textbf{\bibinfo{volume}{A107}},
  \bibinfo{pages}{73} (\bibinfo{year}{1985}).

\bibitem[{\citenamefont{Faci et~al.}(2009)\citenamefont{Faci, Huguet, Queva,
  and Renaud}}]{pconf3}
\bibinfo{author}{\bibfnamefont{S.}~\bibnamefont{Faci}},
  \bibinfo{author}{\bibfnamefont{E.}~\bibnamefont{Huguet}},
  \bibinfo{author}{\bibfnamefont{J.}~\bibnamefont{Queva}}, \bibnamefont{and}
  \bibinfo{author}{\bibfnamefont{J.}~\bibnamefont{Renaud}},
  \bibinfo{journal}{Phys. Rev. D} \textbf{\bibinfo{volume}{80}},
  \bibinfo{pages}{124005} (\bibinfo{year}{2009}), \eprint{0910.1279}.

\bibitem[{\citenamefont{Faci}(2011)}]{pconf5}
\bibinfo{author}{\bibfnamefont{S.}~\bibnamefont{Faci}},
  \bibinfo{journal}{arXiv:1109.2579}  (\bibinfo{year}{2011}).

\bibitem[{\citenamefont{Esposito and Roychowdhury}(2009)}]{Esposito:2008dw}
\bibinfo{author}{\bibfnamefont{G.}~\bibnamefont{Esposito}} \bibnamefont{and}
  \bibinfo{author}{\bibfnamefont{R.}~\bibnamefont{Roychowdhury}},
  \bibinfo{journal}{Int.J.Geom.Meth.Mod.Phys.} \textbf{\bibinfo{volume}{6}},
  \bibinfo{pages}{583} (\bibinfo{year}{2009}), \eprint{0805.1182}.

\bibitem[{\citenamefont{Drew and Gegenberg}(1980)}]{Drew:1980yk}
\bibinfo{author}{\bibfnamefont{M.}~\bibnamefont{Drew}} \bibnamefont{and}
  \bibinfo{author}{\bibfnamefont{J.}~\bibnamefont{Gegenberg}},
  \bibinfo{journal}{Nuovo Cim.} \textbf{\bibinfo{volume}{A60}},
  \bibinfo{pages}{41} (\bibinfo{year}{1980}).

\bibitem[{\citenamefont{Deser and Nepomechie}(1983)}]{Deser:1983tm}
\bibinfo{author}{\bibfnamefont{S.}~\bibnamefont{Deser}} \bibnamefont{and}
  \bibinfo{author}{\bibfnamefont{R.~I.} \bibnamefont{Nepomechie}},
  \bibinfo{journal}{Phys.Lett.} \textbf{\bibinfo{volume}{B132}},
  \bibinfo{pages}{321} (\bibinfo{year}{1983}).

\bibitem[{\citenamefont{Grishchuk and Yudin}(1980)}]{Grishchuk:1980hb}
\bibinfo{author}{\bibfnamefont{L.}~\bibnamefont{Grishchuk}} \bibnamefont{and}
  \bibinfo{author}{\bibfnamefont{V.}~\bibnamefont{Yudin}},
  \bibinfo{journal}{J.Math.Phys.} \textbf{\bibinfo{volume}{21}},
  \bibinfo{pages}{1168} (\bibinfo{year}{1980}).

\bibitem[{\citenamefont{Novello and Neves}(2002)}]{Novello:2002bp}
\bibinfo{author}{\bibfnamefont{M.}~\bibnamefont{Novello}} \bibnamefont{and}
  \bibinfo{author}{\bibfnamefont{R.}~\bibnamefont{Neves}}
  (\bibinfo{year}{2002}), \eprint{gr-qc/0204058}.

\bibitem[{\citenamefont{Constantinescu and
  Crasmareanu}(2009)}]{Constantinescu:2009fk}
\bibinfo{author}{\bibfnamefont{O.}~\bibnamefont{Constantinescu}}
  \bibnamefont{and}
  \bibinfo{author}{\bibfnamefont{M.}~\bibnamefont{Crasmareanu}}
  (\bibinfo{year}{2009}), \eprint{0905.0362}.

\end{thebibliography}
\end{document}